\documentstyle[12pt]{article}
\textheight 9in  \topmargin -.5in   \def\baselinestretch{1.2}
\def\Eq{\begin{equation}}	\def\End{\end{equation}}
\def\frac#1#2{{\textstyle {#1 \over #2}}}
\def\pmb#1{\setbox0=\hbox{#1}\kern-.01em\copy0\kern-\wd0
  \kern.02em\copy0\kern-\wd0\kern-.01em\raise.017em\box0}
\def\Lspace#1{\renewcommand{\baselinestretch}{#1} \large\normalsize}
\def\eg{{\it e.g.}}		\def\d{{\rm d}}
\def\gev{{\rm\,GeV}}		\def\tev{{\rm\,TeV}}
\def\eps{\epsilon}		\def\lam{\lambda}
\def\Zbb{Z^0\to b\bar{b}}	\def\Zcc{Z^0\to c\bar{c}}

\input epsf
\newcounter{fignum}
\def\boxit#1{\vbox{\hrule\hbox{\vrule{#1}\vrule}\hrule}}
\def\Fig#1#2#3#4{\begin{center} \refstepcounter{fignum}  \epsfxsize=#1in
  \label{#2} \hbox to\hsize{\hfil\boxit{\epsffile{#3}}\hfil}
  {\sl Fig.~\thefignum: #4} \end{center}}
\def\putfig#1{Fig.~\ref{#1}}

\begin{document}
\begin{titlepage}
\begin{center}
April, 1996	\hfill CMU-HEP96-05\\
{}~		\hfill DOE/ER/40682-116\\
{}~		\hfill hep-ph/9604325\\
\vskip1.0in {\Large\bf
{~}\pmb{$\Zbb\;$ } Excess from R-Parity Violation}
\vskip.3in
David E. Brahm\footnote{Email: \tt brahm@fermi.phys.cmu.edu}\\
  {\it Carnegie Mellon Physics Dept., Pittsburgh PA 15213}
\end{center}

\vskip.5in
\begin{abstract}
The $\Zbb$ excess, $\Zcc$ deficit, and low left-right asymmetry $A_b$ may be
explained by a single term $\lam C^c B^c B^{\prime c}$ in the superpotential.
This operator violates R-parity and requires a sequential 4th generation.
1-loop diagrams involving squark exchange interfere with the tree-level
processes to give an excess of right-handed $b$ quarks, and a deficit of
right-handed $c$ quarks.  Though the coupling must be large ($\lam\approx 2$ or
$3$), the model is phenomenologically and cosmologically acceptable.
\end{abstract}
\end{titlepage}
\setcounter{footnote}{0}


\section{Some Curious Data}

Though the Standard Model has enjoyed great experimental success, we may be
seeing the first signs of new physics in the decay modes of the $Z^0$ boson.
Combined LEP and SLC data\cite{ewwg} indicate a $3\%$ excess of $\Zbb$ decays
over the Standard Model prediction.  They also hint that the extra $b$'s are
right-handed, and are offset by a corresponding deficit of $c$'s.  The data
are:
\Eq \renewcommand{\arraycolsep}{20pt} \renewcommand{\arraystretch}{1.5}
  \begin{array}{ll}
  R_b \equiv {\Gamma(\Zbb) \over \Gamma(Z^0\to \hbox{hadrons})} = .2219 \pm
    .0017 & (R_b^{SM} = .2156) \\
  R_c \equiv {\Gamma(\Zcc) \over \Gamma(Z^0\to \hbox{hadrons})} = .1540 \pm
    .0074 & (R_c^{SM} = .1724) \\
  R_l \equiv {\Gamma(Z^0\to \hbox{hadrons}) \over \Gamma(Z^0\to l\bar{l})} =
    20.788 \pm  .032 & (R_l^{SM} = 20.786) \\
  A_b \equiv {\Gamma(Z^0\to b_L \bar{b}_L) - \Gamma(Z^0\to b_R \bar{b}_R) \over
    \Gamma(Z^0\to b_L \bar{b}_L) + \Gamma(Z^0\to b_R \bar{b}_R)} = .841 \pm
    .053 & (A_b^{SM} = .935) \\
  A_c = .606 \pm .090 & (A_c^{SM} = .667) \end{array} \End
(The SM prediction for $R_l$ assumes $\alpha_s=.123$ and $m_t = 180\gev$.)  The
discrepancy (``crisis'') in $R_b$ is particularly significant, a $3.7\sigma$
effect.

\section{Our Model, and Its Implications}

We propose a supersymmetric model to explain these data.  It is the MSSM plus
the additional R-parity violating term
\Eq \lam \eps^{abc} C^c_a B^c_b B^{\prime c}_c  \label{newterm}\End
in the superpotential, where $\{a,b,c\}$ are color indices.  $C^c$, $B^c$, and
$B^{\prime c}$ are superfields representing left-handed antiquarks (or
right-handed quarks) and their scalar superpartners.  $C$ is charm, $B$ is
bottom, and $B'$ is the down-type quark in a sequential 4th generation.  The
coupling must be fairly large ($\lam\approx 2$ or $3$).

The MSSM corrections to $R_b$, etc., are known to be insignificant: $\delta R_b
< .002$ \cite{mssm} in the experimentally allowed region of parameter space,
essentially because the sparticles must be heavy and therefore decouple. (See
\cite{haber}, however, for a clever twist on a 4-generation SUSY model).  We
will therefore only calculate corrections from our new term.  We are able to
evade sparticle decoupling by giving $b'$ a mass comparable to the squark
masses.

A sequential 4th generation will give acceptable values of the Peskin-Takeuchi
parameters ($S,T,U$) as long as $t'$ and $b'$ are nearly degenerate.  Our model
does not suffer from the large FCNC's that come with ``exotic'' 4th
generations.  Of course, we need $m_{\nu'} > M_Z/2$.

Our term violates baryon number, but not lepton number.  Thus it cannot induce
proton decay.  Neutron oscillation is highly suppressed by at least 4 loops and
several small CKM angles (since our term does not involve the first
generation), and a factor $\Lambda_{QCD}^5 /\tilde m^4 M_{\tilde g}$ (where
$\tilde m$ is the squark mass and $M_{\tilde g}$ is the gluino mass
\cite{nnbar}); we estimate an effect roughly 7 orders of magnitude weaker than
the experimental limit ($\tau_{n\bar n} > 1.2 \times 10^8 {\rm\,s}$).

Dreiner and Ross \cite{dross} showed that commonly quoted cosmological bounds
\cite{cosmo} can be avoided.  In the presence of our new interaction and of
sphalerons, there are still 3 conserved quantities $(L_1-L_2)$, $(L_1-L_3)$,
and $(L_1-L_4)$ (though $m_{\nu'}$ may break the latter).  A GUT-generated
asymmetry in any of these is preserved.  Near the electroweak phase transition,
sphalerons translate this primordial lepton flavor asymmetry into a baryon
asymmetry.

The large coupling $\lam$ and the 4th-generation Yukawas contribute positively
($\d m^2/\d t > 0$) to the running of scalar masses \cite{chal}.  This effect
is tamed if we have a heavy gluino, \eg
\Eq {\d \tilde m_{B^\prime c}^2 \over \d t} =
  {2\lam^2 \over 8\pi^2} \left[ \tilde m_{B^{\prime c}}^2 \!\!+\! \tilde
    m_{C^c}^2 \!\!+\! \tilde m_{B^c}^2 \!\!+\!\! A^2 \right]
  + {3\lam_{b'}^2 \over 8\pi^2} \left[ \tilde m_{B^{\prime c}}^2 \!\!+\! \tilde
    m_{B'}^2 \!\!+\! m_{h}^2 \!\!+\!\! A^2 \right]
  - {2\over\pi} \left[ {4\alpha_3\over 3} M_3^2
    + {\alpha_1\over 15} M_1^2 \right] \End
where $\lam_{b'} = \sqrt2 m_{b'}/v$, $A$ is the trilinear soft breaking
coefficient, and $M_3$ ($M_1$) is the gluino (bino) mass.  (In SUSY GUT's, the
gaugino masses unify $M_i(M_G)=M_0$, and run like $\alpha_i$, so the gluino is
naturally the heaviest one with $M_3 = 2.9 M_0$.)

The running of $\lam_{b'}$ and $\lam_{t'}$ is discussed in \cite{genu}, where
an upper bound $m_{b'}<156\gev$ is given to keep the couplings perturbative up
to a GUT scale.  The running of $\lam$ is given by
\Eq {\d\lam\over\d t} = \beta(\lam) = {\lam\over 16\pi^2} \left[ 6\lam^2 + 2
  \lam_{b'}^2 - 8 g_3^2 - \frac45 g_1^2 \right] \End
(with $g_1^2 = \frac53 g^{\prime 2}$).  Since we will need $\lam^2(M_Z) > 4.6$,
$\lam$ exhibits a Landau pole at or below $30 M_Z = 2.7\tev$.  Perturbative
unification is thus not possible unless some new physics enters at this scale.

\section{The 1-Loop Diagrams}

The $\Zbb$ excess arises from interference between the tree-level diagram and
the 1-loop diagrams shown in \putfig{diag} (plus 3 others related by $c\!
\leftrightarrow\! b'$, but these are small).  Since only the $B^c$ superfield
enters, only right-handed $b$ production is affected.  The calculation can be
found in \cite[eqs.79,82]{bamert}.  We use the approximation $\{m_{b'}, \tilde
m_{C^c}\} \gg M_Z$, which we find agrees to better than $10\%$ with exact
numerical calculations even for $\tilde m_{C^c} = M_Z$.  In this approximation,
the Standard Model tree-level coupling $g_R^b = s_W^2/3$ is modified by
\Eq \delta g_R^b = {2 |\lam|^2 \over 16\pi^2} (g_R^{b'} - g_L^{b'}) {\cal F}
  \left( m_{b'}^2 \over \tilde m_{C^c}^2 \right), \qquad
  {\cal F}(r) \equiv {r \over (r-1)^2} (r-1-\ln r) \End
${\cal F}(r)$ is positive and monotonically increasing, with ${\cal F}(0)=0$
(satisfying the decoupling theorem as the squark gets heavy), and an asymptotic
value ${\cal F}(\infty) = 1$.

Note that $(g_R^{b'} - g_L^{b'}) = -T_3^{b'} = \frac12$.  The fact that this
has the same sign as $g_R^b = s_W^2/3$ gives an enhancement of $b$ production.
(In any model of this kind, the heavy fermion must have $T_3 < 0$ to give the
right sign for $\delta R_b$.)  We get the right magnitude by setting $\lam^2
{\cal F} = 4.6$, so we need a $\lam \approx 2$ or $3$.

\Fig{5.1}{diag}{zbb_diag.eps}{1-loop diagrams.}

Analogously, the right-handed charm coupling $g_R^c = -2s_W^2/3$ is modified by
\Eq \delta g_R^c = {2 |\lam|^2 \over 16\pi^2} (g_R^{b'} - g_L^{b'}) {\cal F}
  \left( m_{b'}^2 \over \tilde m_{B^c}^2 \right) \End
The magnitude of the charm coupling is reduced, giving a $c$ deficit.

\section{Squark Masses}

If squarks are degenerate, the $c$ deficit is fixed to be twice the $b$ excess.
Choosing only the single parameter $\lam$ would then give
\Eq R_b = .2219 \hbox{ (set)}, \qquad R_c = .1625, \qquad R_l = 20.68, \qquad
  A_b = .89, \qquad A_c = .77 \End
The total hadronic width ($R_l$) is too low (unless $\alpha_s(M_Z) \approx
.15$, which seems unlikely).

Thus we need to take $\tilde m_{B^c} > \tilde m_{C^c}$.  We can adjust the
squark masses to leave the total hadronic width unaltered (so $R_l = R_l^{SM}$
with $\alpha_s(M_Z)=.123$), giving the predictions
\Eq R_b = .2219 \hbox{ (set)}, \qquad R_c = .1656, \qquad A_b = .88, \qquad A_c
  = .73 \End
The value of $A_c$ is still a bit high, but only by $1.4\sigma$.  These results
are in good statistical agreement with all the data.

\section{Some Variations: \pmb{$R_b$}\ Only}

One could treat the $c$ deficit as experimental error, and only explain the
(right-handed) $b$ excess, which under this assumption becomes
\Eq R_b = .2205 \pm .0016 \qquad (R_b^{SM} = .2156) \End
We can do this with a superpotential term $\lam \eps^{abc} T^c_a B^c_b
B^{\prime c}_c$ (to replace eq.~\ref{newterm}) as long as $m_{b'} > m_t$.
\goodbreak

The same result can be achieved with a superpotential term $\lam Q' B^c L'$,
with $Q'=(T',B')$ and $L'=(\nu',\tau')$, if $\tau'$ (or $b'$) is the heaviest
4th generation fermion.  This term has the phenomenological (and cosmological)
advantage of violating only $L_4$, not $B$.

Yet another possibility is $\lam Q_3 B^c L'$, with $Q_3 = (T,B)$.  Then a small
$b_L$ deficit in addition to the $b_R$ excess drives $A_b$ even lower.

\section{Conclusions}

Data indicate an excess of right-handed $b$'s in $Z^0$ decays, offset by a
deficit of $c$'s.  Our model explains these using a single $R_P$-violating term
$\lam C^c B^c B^{\prime c}$ in the superpotential.  Choosing $\lam$
appropriately, and requiring $\tilde m_{B^c} > \tilde m_{C^c}$, we can achieve
agreement with the data to $1.6\sigma$ or better.

We wish to thank Nima Arkani-Hamed, Hsin-Chia Cheng, Lawrence J. Hall, Richard
Holman, Stephen Hsu, and Martin Savage for helpful discussions. This work was
partially supported by the U.S. Dept. of Energy under Contract
DE-FG02-91-ER40682.


\def\pl#1{{\it Phys.\ Lett.} {\bf #1}}
\def\pr#1{{\it Phys.\ Rev.} {\bf #1}}
\def\prl#1{{\it Phys.\ Rev.\ Lett.} {\bf #1}}
\def\np#1{{\it Nucl.\ Phys.} {\bf #1}}
\def\nim#1{{\it Nucl.\ Inst. \& Meth.} {\bf #1}}
\Lspace{.88}

\end{document}